\documentclass[
twocolumn,
]{aastex7}

\usepackage{amsmath}
\usepackage{graphicx}
\usepackage{xcolor}
\usepackage{xparse}
\usepackage{bm}
\usepackage{latexsym}
\usepackage[
capitalize,
noabbrev
]{cleveref}
\usepackage{tikz}
\usepackage[normalem]{ulem}
\usepackage{paralist}
\usepackage{multirow}
\usepackage[yyyymmdd,hhmmss]{datetime}
\usepackage{lipsum}

\NewDocumentCommand{\curl}{o o}{\ensuremath{\IfNoValueTF{#2}{\nabla}{\nabla_{#2}} \times #1}}
\NewDocumentCommand{\grad}{d{_}{_} o}{\ensuremath{\IfNoValueTF{#1}{\nabla}{\nabla_{#1}} #2}}
\NewDocumentCommand{\diverg}{o o}{\ensuremath{\IfNoValueTF{#2}{\nabla}{\nabla_{#2}} \cdot #1}}

\NewDocumentCommand{\years}{o}{\ensuremath{
\IfNoValueF{#1}{#1 \,}
\mathrm{years}
}}

\NewDocumentCommand{\days}{o}{\ensuremath{
\IfNoValueF{#1}{#1 \,}
\mathrm{days}
}}

\NewDocumentCommand{\months}{o}{\ensuremath{
\IfNoValueF{#1}{#1 \,}
\mathrm{months}
}}

\NewDocumentCommand{\s}{o}{\ensuremath{
\IfNoValueF{#1}{#1 \,}
\mathrm{s}
}}

\NewDocumentCommand{\nT}{o}{\ensuremath{
\IfNoValueF{#1}{#1 \,}
\mathrm{nT}
}}

\NewDocumentCommand{\km}{o}{\ensuremath{
\IfNoValueF{#1}{#1 \,}
\mathrm{km}
}}

\NewDocumentCommand{\au}{o}{\ensuremath{
\IfNoValueF{#1}{#1 \,}
\mathrm{AU}
}}

\NewDocumentCommand{\amu}{o}{\ensuremath{
\IfNoValueF{#1}{#1 \,}
\mathrm{amu}
}}

\NewDocumentCommand{\temp}{o}{\ensuremath{
\IfNoValueF{#1}{#1 \times}
\mathrm{10^5 \; K}
}}

\NewDocumentCommand{\pct}{o}{\ensuremath{
\IfNoValueF{#1}{#1 \;}
\%
}}

\NewDocumentCommand{\MK}{o}{\ensuremath{
\IfNoValueF{#1}{#1 \,}
\mathrm{MK}
}}

\NewDocumentCommand{\Rs}{o}{\ensuremath{
\IfNoValueF{#1}{#1 \;}
\mathrm{R_S}
}}

\NewDocumentCommand{\kms}{o}{\ensuremath{
\IfNoValueF{#1}{#1 \;}
\mathrm{km \, s^{-1}}
}}

\NewDocumentCommand{\cc}{o}{\ensuremath{
\IfNoValueF{#1}{#1 \;}
\mathrm{cm^{-3}}
}}

\NewDocumentCommand{\mWcc}{o}{\ensuremath{
\IfNoValueF{#1}{#1 \;}
\mathrm{mW \cc}
}}

\NewDocumentCommand{\mWmsq}{o}{\ensuremath{
\IfNoValueF{#1}{#1 \;}
\mathrm{mW \, m^{-2}}
}}

\NewDocumentCommand{\eV}{o}{\ensuremath{
\IfNoValueF{#1}{#1 \;}
\mathrm{eV}
}}

\NewDocumentCommand{\keV}{o}{\ensuremath{
\IfNoValueF{#1}{#1 \;}
\mathrm{keV}
}}

\NewDocumentCommand{\MeV}{o}{\ensuremath{
\IfNoValueF{#1}{#1 \;}
\mathrm{MeV}
}}

\NewDocumentCommand{\nucleon}{s o}{\ensuremath{
\IfNoValueF{#2}{#2 \;}
\IfBooleanTF{#1}{\mathrm{nucleon}}{\mathrm{nuc}}
}}

\NewDocumentCommand{\MeVnuc}{s o}{\ensuremath{
\IfNoValueF{#2}{#2 \;}
\MeV \! /\IfBooleanTF{#1}{\nucleon*}{\nucleon}
}}

\NewDocumentCommand{\keVe}{o}{\ensuremath{
\IfNoValueF{#1}{#1 \;}
\mathrm{keV/e}
}}

\NewDocumentCommand{\Element}{m}{\ensuremath{\mathrm{#1}}}
\NewDocumentCommand{\QState}{m m}{\ensuremath{\mathrm{#1}^{#2+}}}

\NewDocumentCommand{\Hy}{o}{\IfNoValueTF{#1}{\Element{H}}{\QState{H}{#1}}}
\NewDocumentCommand{\He}{o}{\IfNoValueTF{#1}{\Element{He}}{\QState{He}{#1}}}
\NewDocumentCommand{\C}{o}{\IfNoValueTF{#1}{\Element{C}}{\QState{C}{#1}}}
\NewDocumentCommand{\N}{o}{\IfNoValueTF{#1}{\Element{N}}{\QState{N}{#1}}}
\NewDocumentCommand{\Ox}{o}{\IfNoValueTF{#1}{\Element{O}}{\QState{O}{#1}}}
\NewDocumentCommand{\Ne}{o}{\IfNoValueTF{#1}{\Element{Ne}}{\QState{Ne}{#1}}}
\NewDocumentCommand{\Mg}{o}{\IfNoValueTF{#1}{\Element{Mg}}{\QState{Mg}{#1}}}
\NewDocumentCommand{\Si}{o}{\IfNoValueTF{#1}{\Element{Si}}{\QState{Si}{#1}}}
\NewDocumentCommand{\Su}{o}{\IfNoValueTF{#1}{\Element{S}}{\QState{S}{#1}}}
\NewDocumentCommand{\Ca}{o}{\IfNoValueTF{#1}{\Element{Ca}}{\QState{Ca}{#1}}}
\NewDocumentCommand{\Fe}{o}{\IfNoValueTF{#1}{\Element{Fe}}{\QState{Fe}{#1}}}

\NewDocumentCommand{\FIP}{s o O{=}}{\ensuremath{\mathrm{FIP}
\IfNoValueF{#2}{
#3
\IfBooleanTF{#1}{#2}{\eV[#2]}}
}}

\NewDocumentCommand{\AbSEP}{O{X} O{\Ox}}{\ensuremath{#1/#2}}

\NewDocumentCommand{\PLawExp}{s o}{\ensuremath{b
\IfNoValueF{#2}{\IfBooleanTF{#1}{\approx}{=} #2}}}

\newcommand{\he}{\Element{He}}

\NewDocumentCommand{\MpQ}{o}{\ensuremath{
\IfNoValueTF{#1}{\mathrm{M/Q}}{(\mathrm{M/Q})_{#1}}}}

\NewDocumentCommand{\n}{o o O{=}}{\ensuremath{n
\IfNoValueF{#1}{_{#1}}
\IfNoValueF{#2}{
\IfNoValueTF{#3}{=}{#3} \cc[#2]
}
}}

\NewDocumentCommand{\dn}{o o O{=}}{\ensuremath{\delta n
\IfNoValueF{#1}{_{#1}}
\IfNoValueF{#2}{
\IfNoValueTF{#3}{=}{#3} \cc[#2]
}
}}

\NewDocumentCommand{\dnn}{s o o O{=}}{\ensuremath{
\IfBooleanTF{#1}{\dn[#2] / \n[#2]}{\abs{\dn[#2] / \n[#2]}}
\IfNoValueF{#3}{
\IfNoValueTF{#4}{=}{#4} {#3}
}
}}

\NewDocumentCommand{\m}{o o O{=}}{\ensuremath{\rho
\IfNoValueF{#1}{_{#1}}
\IfNoValueF{#2}{
\IfNoValueTF{#3}{=}{#3} \cc[#2]
}
}}

\NewDocumentCommand{\W}{d<> o O{=}}{\ensuremath{W
\IfNoValueF{#1}{_{#1}}
\IfNoValueF{#2}{
\IfNoValueTF{#3}{=}{#3} \mWmsq[#2]
}
}}

\NewDocumentCommand{\Wg}{d<> o O{=}}{\ensuremath{W
\IfNoValueTF{#1}{_g}{_{g,#1}}
\IfNoValueF{#2}{
\IfNoValueTF{#3}{=}{#3} \mWmsq[#2]
}
}}

\NewDocumentCommand{\Wk}{d<> o O{=}}{\ensuremath{W
\IfNoValueTF{#1}{_K}{_{K,#1}}
\IfNoValueF{#2}{
\IfNoValueTF{#3}{=}{#3} \mWmsq[#2]
}
}}

\NewDocumentCommand{\WA}{d<> o O{=}}{\ensuremath{W
\IfNoValueTF{#1}{_A}{_{A,#1}}
\IfNoValueF{#2}{
\IfNoValueTF{#3}{=}{#3} \mWmsq[#2]
}
}}

\NewDocumentCommand{\We}{d<> o O{=}}{\ensuremath{W
\IfNoValueTF{#1}{_E}{_{E,#1}}
\IfNoValueF{#2}{
\IfNoValueTF{#3}{=}{#3} \mWmsq[#2]
}
}}

\NewDocumentCommand{\vsw}{s o O{=}}{\ensuremath{v_\sw
\IfNoValueF{#2}{
\IfNoValueTF{#3}{=}{#3}
\IfBooleanTF{#1}{#2}{\kms[#2]}}
}}

\NewDocumentCommand{\vs}{s o O{=}}{\ensuremath{v_s
\IfNoValueF{#2}{
\IfNoValueTF{#3}{=}{#3}
\IfBooleanTF{#1}{#2}{\kms[#2]}}
}}

\NewDocumentCommand{\vsigma}{s o O{=}}{\ensuremath{v_\sigma
\IfNoValueF{#2}{
\IfNoValueTF{#3}{=}{#3}
\IfBooleanTF{#1}{#2}{\kms[#2]}}
}}

\NewDocumentCommand{\vAhe}{s o O{=}}{\ensuremath{v_s\!\left(\ahe\right)
\IfNoValueF{#2}{
\IfNoValueTF{#3}{=}{#3}
\IfBooleanTF{#1}{#2}{\kms[#2]}}
}}

\NewDocumentCommand{\vTbar}{s o O{=} d<>}{\ensuremath{v_s\!\left(\IfNoValueTF{#4}{\Tbar}{\Tbar<#4>}\right)
\IfNoValueF{#2}{
\IfNoValueTF{#3}{=}{#3}
\IfBooleanTF{#1}{#2}{\kms[#2]}}
}}

\NewDocumentCommand{\vslow}{s o O{=}}{\ensuremath{v_\mathrm{slow}
\IfNoValueF{#2}{
\IfNoValueTF{#3}{=}{#3}
\IfBooleanTF{#1}{#2}{\kms[#2]}}
}}

\NewDocumentCommand{\vfast}{s o O{=}}{\ensuremath{v_\mathrm{fast}
\IfNoValueF{#2}{
\IfNoValueTF{#3}{=}{#3}
\IfBooleanTF{#1}{#2}{\kms[#2]}}
}}

\NewDocumentCommand{\valpha}{s o O{=}}{\ensuremath{v_n
\IfNoValueF{#2}{
\IfNoValueTF{#3}{=}{#3}
\IfBooleanTF{#1}{#2}{\kms[#2]}}
}}

\NewDocumentCommand{\ve}{s o O{=}}{\ensuremath{v_{E}
\IfNoValueF{#2}{
\IfNoValueTF{#3}{=}{#3}
\IfBooleanTF{#1}{#2}{\kms[#2]}}
}}

\NewDocumentCommand{\vWk}{s o O{=}}{\ensuremath{v_K
\IfNoValueF{#2}{
\IfNoValueTF{#3}{=}{#3}
\IfBooleanTF{#1}{#2}{\kms[#2]}}
}}

\NewDocumentCommand{\vi}{s o O{=}}{\ensuremath{v_i
\IfNoValueF{#2}{
\IfNoValueTF{#3}{=}{#3}
\IfBooleanTF{#1}{#2}{\kms[#2]}}
}}

\NewDocumentCommand{\vIP}{s o O{=}}{\ensuremath{v_{IP}
\IfNoValueF{#2}{
\IfNoValueTF{#3}{=}{#3}
\IfBooleanTF{#1}{#2}{\kms[#2]}}
}}

\NewDocumentCommand{\vIPW}{s o O{=}}{\ensuremath{v_{IPW}
\IfNoValueF{#2}{
\IfNoValueTF{#3}{=}{#3}
\IfBooleanTF{#1}{#2}{\kms[#2]}}
}}

\NewDocumentCommand{\vel}{s d<> o O{=}}{\ensuremath{v\IfNoValueF{#2}{_{#2}}
\IfNoValueF{#3}{
#4
\IfBooleanTF{#1}{#3}{\kms[#3]}}
}}

\NewDocumentCommand{\vv}{s o O{=}}{\ensuremath{v_v
\IfNoValueF{#2}{
\IfNoValueTF{#3}{=}{#3}
\IfBooleanTF{#1}{#2}{\kms[#2]}}
}}

\NewDocumentCommand{\vn}{s o O{=}}{\ensuremath{v_n
\IfNoValueF{#2}{
\IfNoValueTF{#3}{=}{#3}
\IfBooleanTF{#1}{#2}{\kms[#2]}}
}}

\NewDocumentCommand{\As}{s o O{=}}{\ensuremath{A_s
\IfNoValueF{#2}{
\IfNoValueTF{#3}{=}{#3}
\IfBooleanTF{#1}{#2}{#2 \%}}
}}

\NewDocumentCommand{\Ra}{d<> o O{=}}{\ensuremath{r
\IfNoValueTF{#1}{_A}{_{A,#1}}
\IfNoValueF{#2}{
\IfNoValueTF{#3}{=}{#3} R_{\bigodot}
}
}}

\NewDocumentCommand{\grate}{o o}{\ensuremath{
\gamma\IfNoValueF{#1}{/\Omega_{#1}}
\IfNoValueF{#2}{= 10^{{#2}}}
}}

\NewDocumentCommand{\gmax}{o}{\ensuremath{
\gamma_\mathrm{max}\IfNoValueF{#1}{/\Omega_{#1}}
}}

\NewDocumentCommand{\kvec}{o}{\ensuremath{
\vec{k} \rho\IfNoValueF{#1}{{_{#1}}}
}}

\NewDocumentCommand{\kpar}{o}{\ensuremath{
{k_\parallel} \rho\IfNoValueF{#1}{{_{#1}}}
}}

\NewDocumentCommand{\kper}{o}{\ensuremath{
{k_\perp} \rho\IfNoValueF{#1}{{_{#1}}}
}}

\NewDocumentCommand{\ani}{s o}{\ensuremath{
R\IfNoValueF{#2}{_{#2}}
\IfBooleanT{#1}{\, [\perp\!/\!\parallel]}
}}

\NewDocumentCommand{\Temp}{s d{_}{_} o O{=}}{\ensuremath{T{\IfNoValueF{#2}{_{#2}}}
\IfNoValueF{#3}{{#4}\IfBooleanTF{#1}{#3}{\MK[#3]}}
}}

\NewDocumentCommand{\Trat}{s o o o}{\ensuremath{
T_{\IfNoValueF{#4}{{#4};}#2}/T_{\IfNoValueF{#4}{{#4};}#3}
 \IfBooleanT{#1}{\, [\#]}
}}

\NewDocumentCommand{\TratHeH}{s o}{\ensuremath{
T_{\IfNoValueF{#2}{{#2};}\He}/T_{\IfNoValueF{#2}{{#2};}\Hy}
 \IfBooleanT{#1}{\, [\#]}
}}

\NewDocumentCommand{\Tbar}{s d<> o O{=}}{\ensuremath{
\bar{T}{\IfNoValueF{#2}{_{#2}}}
\IfNoValueF{#3}{#4 #3}
 \IfBooleanT{#1}{\, [\#]}
}}

\NewDocumentCommand{\Tbars}{s o O{=}}{\ensuremath{
\bar{T}{_s}
\IfNoValueF{#2}{#3 #2}
 \IfBooleanT{#1}{\, [\#]}
}}

\NewDocumentCommand{\pbeta}{s o}{\ensuremath{
\beta\IfNoValueF{#2}{_{#2}}
 \IfBooleanT{#1}{\, [\#]}
}}

\NewDocumentCommand{\pbetaR}{o}{\ensuremath{
(\pbeta[\parallel
\IfNoValueF{#1}{;#1}], \ani[#1])
}}

\NewDocumentCommand{\dv}{o}{\ensuremath{\Delta v\IfNoValueF{#1}{_{#1}}}}
\NewDocumentCommand{\ca}{o}{\ensuremath{C_{A\IfNoValueF{#1}{;#1}}}}
\NewDocumentCommand{\dvca}{o o}{\ensuremath{\dv[#1]/\ca[#2]}}

\NewDocumentCommand{\nuc}{o}{\ensuremath{\nu_{c\IfNoValueF{#1}{;#1}}}}
\NewDocumentCommand{\Nc}{o}{\ensuremath{N_{c\IfNoValueF{#1}{;#1}}}}
\NewDocumentCommand{\Ac}{o}{\ensuremath{A_{c\IfNoValueF{#1}{;#1}}}}

\NewDocumentCommand{\tauEXP}{o}{\ensuremath{
\tau_{\mathrm{exp}\IfNoValueF{#1}{;#1}
}}}

\NewDocumentCommand{\tauCC}{o}{\ensuremath{
\tau_{\mathrm{C}\IfNoValueF{#1}{;#1}
}}}

\NewDocumentCommand{\SSN}{o}{\ensuremath{\mathrm{SSN}
\IfNoValueF{#1}{#1}}}
\NewDocumentCommand{\NSSN}{o}{\ensuremath{\mathrm{NSSN}
\IfNoValueF{#1}{#1}}}

\newcommand{\sw}{\ensuremath{\mathrm{sw}}}

\NewDocumentCommand{\qpar}{o}{\ensuremath{
q_{\parallel
\IfNoValueF{#1}{;#1}
}}}

\NewDocumentCommand{\edv}{o}{\ensuremath{
\tilde{E}_{\dv[#1]
}}}

\NewDocumentCommand{\ndays}{o}{
\ensuremath{N_\mathrm{days}{\IfNoValueF{#1}{= {#1}}}}
}

\NewDocumentCommand{\se}{o}{\ensuremath{
S{\IfNoValueF{#1}{_{#1}}}
}}

\NewDocumentCommand{\ab}{o}{\ensuremath{
A{\IfNoValueF{#1}{_{#1}}}
}}

\NewDocumentCommand{\ahe}{s o O{=}}{\ensuremath{\ab[\he]
\IfNoValueF{#2}{
\IfNoValueTF{#3}{=}{#3}
\IfBooleanTF{#1}{#2}{#2 \%}}
}}

\NewDocumentCommand{\corr}{o}{\ensuremath{
\rho
\IfNoValueF{#1}{(#1)}
}}

\NewDocumentCommand{\xhel}{s o O{=} o}{\ensuremath{
\IfBooleanTF{#1}{\sigma_{c\IfNoValueF{#4}{,#4}}}{\abs{\sigma_{c\IfNoValueF{#4}{,#4}}}}
\IfNoValueF{#2}{
\IfNoValueTF{#3}{=}{#3}
#2}
}}

\NewDocumentCommand{\SpecInd}{o}{\ensuremath{\gamma
\IfNoValueF{#1}{_{#1}}}}
\NewDocumentCommand{\QT}{o}{\ensuremath{\mathrm{QT}
\IfNoValueF{#1}{= #1}}}

\NewDocumentCommand{\pten}{o m}{\ensuremath{
\IfNoValueF{#1}{#1 \times}10^{#2}
}}

\NewDocumentCommand{\abs}{m}{\ensuremath{\left| #1 \right|}}

\NewDocumentCommand{\fcn}{m m o O{=}}{\ensuremath{#1\left(#2\right)\IfNoValueF{#3}{#4 #3}}
}

\NewDocumentCommand{\nsigma}{o}{\ensuremath{\IfNoValueF{#1}{#1}{\relax}\sigma}
}

\newcommand{\citepossessive}[1]{\citeauthor{#1}'s (\citeyear{#1})}

\definecolor{C0}{HTML}{1f77b4}
\definecolor{C1}{HTML}{ff7f0e}
\definecolor{C2}{HTML}{2ca02c}
\definecolor{C3}{HTML}{d62728}
\definecolor{C4}{HTML}{9467bd}
\definecolor{C5}{HTML}{8c564b}

\definecolor{QTFitGreen}{HTML}{2ca02c}
\definecolor{DodgerBlue}{HTML}{1e90ff}
\definecolor{Fuchsia}{HTML}{ff00ff}
\definecolor{TabGreen}{HTML}{2ca02c}
\definecolor{Cyan}{HTML}{00ffff}
\definecolor{LimeGreen}{HTML}{32cd32}
\definecolor{Lime}{HTML}{00ff00}

\definecolor{MaxPink}{HTML}{e377c2}
\definecolor{MinPurple}{HTML}{9467bd}

\definecolor{q}{HTML}{228B22}
\definecolor{wc}{HTML}{FF8C00}
\definecolor{dnc}{HTML}{FF00FF}
\definecolor{todo}{HTML}{e13748}
\definecolor{ben}{HTML}{e13748}
\definecolor{bob}{HTML}{0080FF}

\NewDocumentCommand{\question}{s o m}{\IfBooleanF{#1}{\textcolor{q}{\textbf{Q}\IfNoValueF{#2}{ (#2)}: \textit{#3}}}}
\NewDocumentCommand{\answer}{s o m}{\IfBooleanF{#1}{\textcolor{q}{\textbf{A}\IfNoValueF{#2}{ (#2)}: \textit{#3}}}}

\NewDocumentCommand{\wc}{s m}{\IfBooleanTF{#1}{#2}{\textcolor{wc}{\textbf{WC:} \textit{#2}}}}
\NewDocumentCommand{\ws}{s m}{\IfBooleanTF{#1}{#2}{\textcolor{wc}{\textbf{WS:} \textit{#2}}}}

\NewDocumentCommand{\delete}{s m}{\IfBooleanF{#1}{\textcolor{todo}{\textbf{Delete:} \textit{#2}}}}
\NewDocumentCommand{\todo}{s o m}{\IfBooleanF{#1}{\textcolor{todo}{\textbf{TODO}\IfNoValueF{#2}{ (#2)}: \textit{#3}}}}
\NewDocumentCommand{\verify}{s o m}{\IfBooleanTF{#1}{#3}{\textcolor{todo}{\textbf{VERIFY}\IfNoValueF{#2}{ (#2)}: \textit{#3}}}}
\NewDocumentCommand{\goal}{s o m}{\IfBooleanTF{#1}{#3}{\textcolor{todo}{\textbf{GOAL}\IfNoValueF{#2}{ (#2)}: \textit{#3}}}}

\NewDocumentCommand{\move}{s o m}{\textcolor{dnc}{\textbf{\IfBooleanTF{#1}{Duplicate}{Move}}\IfNoValueF{#2}{ (#2)}: \textit{#3}}}
\NewDocumentCommand{\dupe}{o m}{\move*[#1]{#2}}
\NewDocumentCommand{\intro}{s m}{\IfBooleanTF{#1}{\dupe[Intro]{#2}}{\move[Intro]{#2}}}
\NewDocumentCommand{\dnc}{s m}{\IfBooleanTF{#1}{\dupe[DnC]{#2}}{\move[DnC]{#2}}}
\NewDocumentCommand{\fw}{s m}{\IfBooleanTF{#1}{\dupe[Future Work]{#2}}{\move[DnC]{#2}}}

\DeclareDocumentCommand{\EmptyTimes}{O{black}}{\ensuremath{\mathord{\begin{tikzpicture}[line width=0.2ex, x=1.5ex, y=1.5ex]
\draw[color=#1] (0, 0.25) -- (0.25, 0.5) -- (0, 0.75) -- (0.25, 1.0) -- (0.5, 0.75) -- (0.75, 1.0) -- (1.0, 0.75) -- (0.75, 0.5) -- (1.0, 0.25) -- (0.75, 0) -- (0.5, 0.25) -- (0.25, 0) -- cycle;
\end{tikzpicture}}}}

\DeclareDocumentCommand{\SolidBand}{O{black} D{<}{>}{1}}{\ensuremath{\mathord{\begin{tikzpicture}[line width=1.25ex, x=1.25ex, y=1.25ex, yshift=5ex]
\draw[color=#1, opacity=#2] (0,0.5) -- (1.5,0.5);
\draw[opacity=0, line width=0.1ex] (0,0) -- (1.5,0);
\end{tikzpicture}}}}

\DeclareDocumentCommand{\SolidBandVertLines}{O{black} D{<}{>}{1}}{\ensuremath{\mathord{\begin{tikzpicture}[line width=1.25ex, x=1.25ex, y=1.25ex, yshift=5ex]
\draw[color=#1, opacity=#2] (0,0.5) -- (1.5,0.5);
\draw[opacity=0, line width=0.1ex] (0,0) -- (1.5,0);
\draw[color=#1, line width=0.2ex] (0, 0) -- (0, 1);
\draw[color=#1, line width=0.2ex] (1.5, 0) -- (1.5, 1);
\end{tikzpicture}}}}

\DeclareDocumentCommand{\SolidLine}{O{black}}{\ensuremath{\mathord{\begin{tikzpicture}[line width=0.3ex, x=1.25ex, y=1.25ex, yshift=5ex]
\draw[color=#1] (0,0.5) -- (1,0.5);
\draw[opacity=0] (0,0) -- (1,0);
\end{tikzpicture}}}}

\DeclareDocumentCommand{\DashedLine}{O{black} o D{<}{>}{1}}{\ensuremath{\mathord{\begin{tikzpicture}[line width=0.3ex, x=1.25ex, y=1.25ex, yshift=5ex]
\IfNoValueF{#2}{\draw[color=#2, opacity=#3] (0, 0.5) -- (1.75, 0.5);}
\draw[color=#1, opacity=#3] (0,0.5) -- (0.75,0.5);
\draw[color=#1, opacity=#3] (1.0,0.5) -- (1.75,0.5);
\draw[opacity=0] (0,0) -- (1.25,0);
\end{tikzpicture}}}}

\DeclareDocumentCommand{\HighlightDashedLine}{O{black} O{LimeGreen} D{<}{>}{1}}{\ensuremath{\mathord{\begin{tikzpicture}[line width=0.3ex, x=1.25ex, y=1.25ex, yshift=5ex]
\draw[color=#2, opacity=#3, line width=0.8ex] (0, 0.5) -- (1.75, 0.5);
\draw[color=#1] (0,0.5) -- (0.75,0.5);
\draw[color=#1] (1.0,0.5) -- (1.75,0.5);
\draw[opacity=0] (0,0) -- (1.25,0);
\end{tikzpicture}}}}

\DeclareDocumentCommand{\DotDotDotLine}{O{white} O{black}}{\ensuremath{\mathord{\begin{tikzpicture}[line width=0.3ex, x=1.25ex, y=1.25ex, yshift=5ex]
\draw[color=#1] (0, 0.5) -- (1.5, 0.5);
\draw[color=#2] (0,0.5) -- (0.3,0.5);
\draw [color=#2](0.6,0.5) -- (0.9,0.5);
\draw[color=#2] (1.2,0.5) -- (1.5,0.5);
\draw[opacity=0] (0,0) -- (1.5,0);
\end{tikzpicture}}}}

\DeclareDocumentCommand{\DotDotDotLineTwo}{O{yellow} O{black}}{\ensuremath{\mathord{\begin{tikzpicture}[line width=0.3ex, x=1.25ex, y=1.25ex, yshift=5ex]
\draw[color=#1] (0,0.5) -- (1.5,0.5);
\draw[color=#2] (0,0.5) -- (0.3,0.5);
\draw[color=#2] (0.6,0.5) -- (0.9,0.5);
\draw[color=#2] (1.2,0.5) -- (1.5,0.5);
\draw[opacity=0] (0,0) -- (1.5,0);
\end{tikzpicture}}}}

\DeclareDocumentCommand{\DashDotDotDotLine}{O{white} O{black}}{\ensuremath{\mathord{\begin{tikzpicture}[line width=0.3ex, x=1.25ex, y=1.25ex, yshift=5ex]
\draw[color=#1] (0, 0.5) -- (2.8, 0.5);
\draw[color=#2] (0,0.5) -- (1,0.5);
\draw[color=#2] (1.3,0.5) -- (1.6,0.5);
\draw[color=#2] (1.9,0.5) -- (2.2,0.5);
\draw[color=#2] (2.5,0.5) -- (2.8,0.5);
\draw[opacity=0] (0,0) -- (2.8,0);
\end{tikzpicture}}}}

\DeclareDocumentCommand{\Circle}{O{black}}{\ensuremath{\mathord{\begin{tikzpicture}[line width=0.3ex, x=1.25ex, y=1.25ex, yshift=5ex]
\draw[color=#1] circle (0.75ex);
\end{tikzpicture}}}}

\NewDocumentCommand{\sect}{o m}{Section~\ref{sec:#2}\IfNoValueF{#1}{ #1}}
\NewDocumentCommand{\eq}{o m}{\cref{eq:#2}\IfNoValueF{#1}{ #1}}
\NewDocumentCommand{\tbl}{o m}{\cref{tbl:#2}\IfNoValueF{#1}{ #1}}

\NewDocumentCommand{\BlindText}{O{6}}{\todo{Remove blind text. This is here to help figures render nicely.}
\textcolor{white}{\lipsum*[1-#1]}} \usepackage{multirow}
\usepackage{tablefootnote}

\newcommand{\figwidth}{\linewidth}
\makeatletter
\@ifclasswith{aastex631}{modern}{\renewcommand{\figwidth}{0.3\linewidth}}{\relax}
\makeatother

\NewDocumentCommand{\PlotHist}{s}{
\IfBooleanTF{#1}{\begin{figure*}}{\begin{figure}}
\begin{centering}
\includegraphics[width=\figwidth]{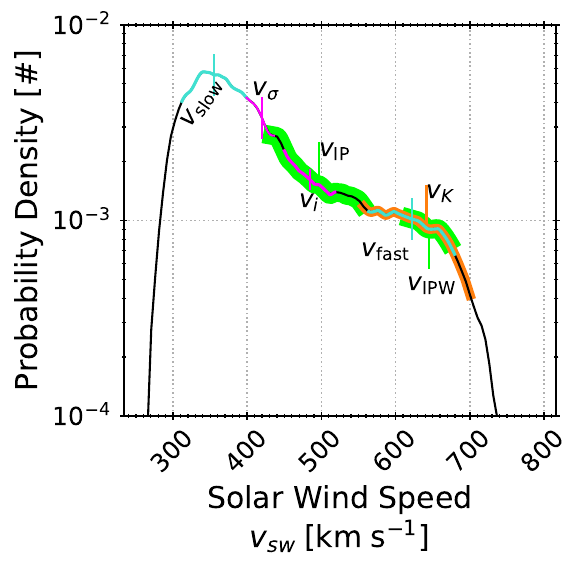}
\end{centering}
\caption{\label{fig:hist}
Probability density function (PDF) of the solar wind speed during the declining phase of solar activity.
The turquoise regions are the slow and fast wind peaks, labeled \vslow\ and \vfast.
The pink regions indicate \vsigma\ and \vi.
\vsigma\ is the range of speeds over which \grad_\vsw_[\ahe] decreases as the Alfvénicity increases.
\vi\ is the range of speeds over which Gaussians fit to the slow and fast wind distributions intersect.
The range of speeds near-Earth predicted from observations of the near-Sun kinetic energy flux is highlighted in orange and labeled \vWk. Green regions indicate the speeds derived using the isopoly model without (\vIP) and with (\vIPW) Alfvén wave forcing.
\cref{tbl:v} summarizes these values.
}
\IfBooleanTF{#1}{\end{figure*}}{\end{figure}}
}

\NewDocumentCommand{\PlotKeFlux}{s}{
\IfBooleanTF{#1}{\begin{figure*}}{\begin{figure}}
\begin{centering}
\includegraphics[width=\figwidth]{vsw-West-LogNormals-SolMin}
\end{centering}
\caption{\label{fig:KE}
The total observed kinetic energy flux as a function of solar wind speed as observed near Earth during solar minima.
This region is plotted in magenta and labeled \emph{Observed}.
The speed range highlighted in turquoise corresponds to the fast wind peak of the distribution of near-Earth observations collected during solar minima.
The purple region labeled \emph{Predicted} is the range of kinetic energy fluxes predicted near Earth from Parker Solar Probe observations of the radial scaling of the total and kinetic energy fluxes during Encounters 1, 2, 4, and 5.
}
\IfBooleanTF{#1}{\end{figure*}}{\end{figure}}
}

\NewDocumentCommand{\PlotKeFluxHeatMap}{s}{
\IfBooleanTF{#1}{\begin{figure*}}{\begin{figure}}
\begin{centering}
\includegraphics[width=\figwidth]{vsw-Wearth-extraction}
\end{centering}
\caption{\label{fig:KE:H2D}
A 2D heat map showing the frequency of observations as a function of solar wind speed and kinetic energy flux observed near Earth.
Because slow solar wind is observed more frequently than fast solar wind (\cref{fig:hist}), each column in the heat map is normalized to its maximum value to remove this sampling artifact.
The black shaded region identifies log-normal distributions fit to each column.
}
\IfBooleanTF{#1}{\end{figure*}}{\end{figure}}
}

\NewDocumentCommand{\PlotKeFluxCombined}{s}{
\IfBooleanTF{#1}{\begin{figure*}}{\begin{figure}}
\begin{centering}
\includegraphics[width=\figwidth]{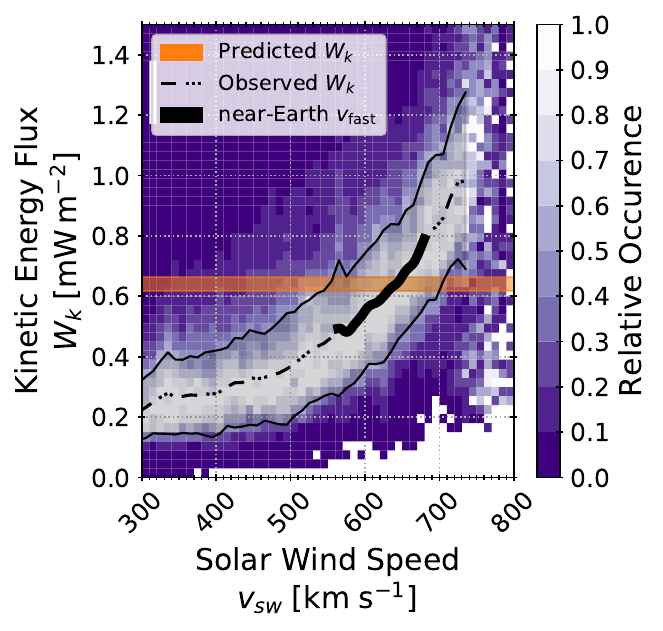}
\end{centering}
\caption{\label{fig:KE}
A 2D heat map showing the frequency of observations as a function of solar wind speed and kinetic energy flux observed near Earth.
Because slow solar wind is observed more frequently than fast solar wind (\cref{fig:hist}), each column in the heat map is normalized to its maximum value to remove this sampling artifact.
The black dash-dotted line is the central value in each column; the black solid lines are the spread of values.
The region highlighted with a thick black line corresponds to \vfast.
The orange horizontal band is the range of kinetic energy fluxes predicted near Earth from Probe observations of the radial scaling of the total and kinetic energy fluxes during Encounters 1, 2, 4, and 5.
}
\IfBooleanTF{#1}{\end{figure*}}{\end{figure}}
}

\NewDocumentCommand{\PlotLogNormalFit}{s}{
\IfBooleanTF{#1}{\begin{figure*}}{\begin{figure}}
\begin{centering}
\includegraphics[page=38,width=\figwidth, trim=0 0 6cm 5ex, clip]{KeFlux-1D-Trend-Hists}
\end{centering}
\caption{\label{fig:log-normal-fit}
An example of the log-normal distribution fit to observations of solar wind speeds observed near Earth in the range of speeds 620 to \kms[630].
The binned observations are in black.
The fit is plotted in a dash-dotted orange line.
}
\IfBooleanTF{#1}{\end{figure*}}{\end{figure}}
}

\NewDocumentCommand{\PlotHistFits}{s}{
\IfBooleanTF{#1}{\begin{figure*}}{\begin{figure}}
\begin{centering}
\includegraphics[width=\figwidth]{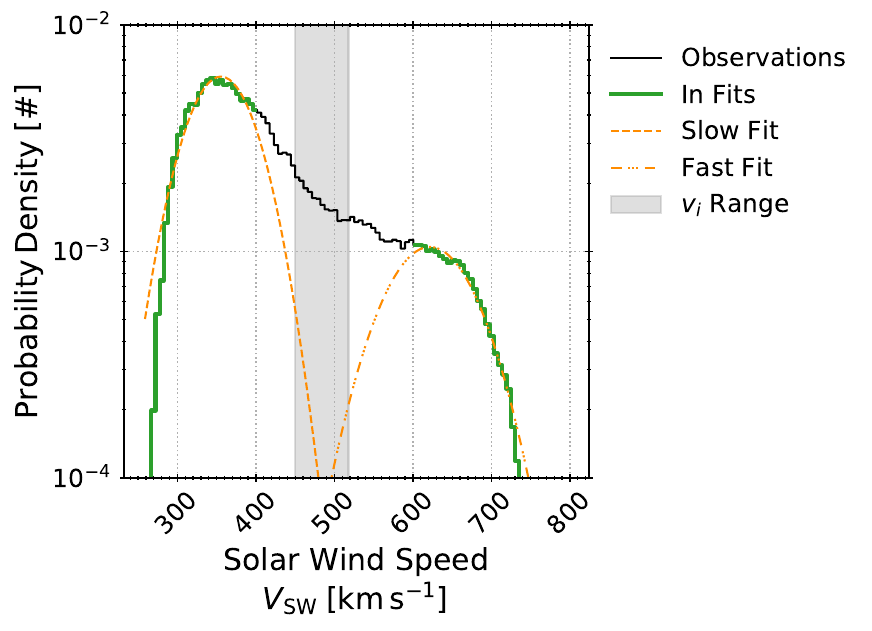}
\end{centering}
\caption{\label{fig:hist:fits}
The PDF from \cref{fig:hist} highlighting the observations used in the Gaussian fits to slow and fast wind in green and the resulting fits in orange.
The gray region indicates the speed range of overlap \vi\ between the fits.
}
\IfBooleanTF{#1}{\end{figure*}}{\end{figure}}
}

\newcommand{\eqWk}{
\begin{equation}\label{eq:Wk}
\Wk<s> = \frac{1}{2}\m[s] \vel<s>^3
\end{equation}
}

\newcommand{\eqRadialScaling}{
\begin{equation}\label{eq:scaling}
f(R) = c \left(\frac{R}{R_{S}}\right)^b.
\end{equation}
}

\NewDocumentCommand{\TblFit}{s}{
\IfBooleanTF{#1}{\begin{table*}}{\begin{table}}
\centering
\begin{tabular}{cccc}
\multicolumn{4}{c}{\textbf{Gaussian Fit Parameters for}}\\
\multicolumn{4}{c}{\textbf{Solar Wind Speed Distributions}}\\
\hline
\hline
 Parameter    &  Units  & Slow Peak & Fast Peak \\
\hline
Center ($\mu$) & $\left[\kms\right]$ &  $355 \pm 1$ & $622 \pm 2$ \\
Width   ($\sigma$)  & $\left[\kms\right]$  &    $44 \pm 2$   &  $58 \pm 2$    \\
Amplitude ($A$) & $\left[10^{-3}\right]$ &  $ 5.9 \pm 0.1$ &  $1.05 \pm 0.01$\\
\hline
\hline
\end{tabular}
\caption{\label{tbl:fits}
Parameters characterizing the Gaussians fit to the slow and fast wind peaks in \cref{fig:hist}.
}
\IfBooleanTF{#1}{\end{table*}}{\end{table}}
}

\NewDocumentCommand{\TblScaling}{s}{
\IfBooleanTF{#1}{\begin{table*}}{\begin{table}}
\centering
\begin{tabular}{cccc}
\multicolumn{4}{c}{\textbf{Radial Scaling Parameters for the}}\\
\multicolumn{4}{c}{\textbf{Solar Wind Energy Flux}}\\
\hline
\hline
Energy Flux   & Symbol & Coefficient $c$ & Exponent $b$ \\
\hline
Total & \W & $52.1 \pm 1.4$ &  $-1.92 \pm 0.007$ \\
KE Fraction & $\Wk/\W$ & $0.04 \pm 0.001$  &    $0.41 \pm 0.007$ \\
\hline
\hline
\end{tabular}
\caption{\label{tbl:scaling}
Parameters from the radial scalings in \cref{eq:scaling} for the total energy flux and kinetic energy flux fraction from \citet[their Fig.~6]{Liu2021c}.
}
\IfBooleanTF{#1}{\end{table*}}{\end{table}}
}

\NewDocumentCommand{\TblVel}{s}{
\IfBooleanTF{#1}{\begin{table*}}{\begin{table}}
\centering
\begin{tabular}{cc l}
\multicolumn{3}{c}{\textbf{Summary of Speed Ranges}} \\
\hline
\hline
Symbol & Speed Range & Description  \\
\hline
$v_\mathrm{IPW}$ &  $620$ to $671$ & Near-Earth speed predicted \\
& & from the isopoly model with  \\
& & Alfvén wave forcing. \\
$v_\mathrm{fast}$ &  $622 \pm 58$ & Peak of the fast wind distribut-  \\
& & ion observed near-Earth. \\
$v_K$ &  $557$ to $700$ & Near-Earth speed predicted \\
& & from near-Sun kinetic energy \\
& &  flux observed by Probe. \\
$v_\mathrm{IP}$ &  $434$ to $554$ & Near-Earth speed predicted \\
& &  from the isopoly model with- \\
& & out Alfvén wave forcing.\\
$v_i$ &  $484 \pm 34$ & Speeds at which Gaussians fit \\
& &  to the fast and slow wind dis- \\
& & tributions overlap.\\
$\vsigma$ &  $399$ to $436$ & Speed range identified by  \\
& & \citet{Wind:SWE:ahe:dnn}  \\
& & as the transition between fast  \\
& & and slow wind, accounting for  \\
& & the helium abundance, normal- \\
& & ized cross helicity, and solar wind \\
& &  compressibility. \\
$v_\mathrm{slow}$ &  $355 \pm 44$ & Peak of the slow wind distri- \\
& & bution observed near-Earth. \\
\hline
\hline
\end{tabular}
\caption{\label{tbl:v}
Ranges of speeds derived and referenced in this work.
All units are \kms.
}
\IfBooleanTF{#1}{\end{table*}}{\end{table}}
} 
\graphicspath{{./}{figures/}}
\DeclareGraphicsExtensions{.pdf,.png,.jpg,.eps,}

\received{\today}
\submitjournal{ApJ Letters}

\NewDocumentCommand{\gradAhe}{o}{\ensuremath{
\IfNoValueTF{#1}{\grad_\vsw_[\ahe]}{\grad_\vsw_[\ahe][#1]}}}

\begin{document}

\title{Characterizing the Impact of Alfvén Wave Forcing in Interplanetary Space on the Distribution of near-Earth Solar Wind Speeds}

\newcommand{\GSFC}{
\affiliation{Heliophysics Science Division, NASA, Goddard Space Flight Center, Greenbelt, MD 20771, USA}
}

\author{B.\ L.\ Alterman}
\email{b.l.alterman@nasa.gov}
\GSFC

\begin{abstract}

Broadly, solar wind source regions can be classified by their magnetic topology as intermittently and continuously open to the heliosphere.
Early models of solar wind acceleration do not account for the fastest, non-transient solar wind speeds observed near-Earth and energy must be deposited into the solar wind after it leaves the Sun.
Alfvén wave energy deposition and thermal pressure gradients are likely candidates and the relative contribution of each acceleration mechanism likely depends on the source region.
Although solar wind speed is a rough proxy for solar wind source region, it cannot unambiguously identify source region topology.

Using near-Sun observations of the solar wind's kinetic energy flux, we predict the expected kinetic energy flux near-Earth.
This predicted kinetic energy flux corresponds to the range of solar wind speeds observed in the fast solar wind and infer that the solar wind's near-Sun kinetic energy flux is sufficient to predict the distribution of fastest, non-transient speeds observed near Earth.
Applying a recently developed model of solar wind evolution in the inner heliosphere, we suggest that the acceleration required to generate this distribution of fastest, non-transient speeds is likely due to the continuous deposition of energy by Alfvén wave forcing during the solar wind's propagation through interplanetary space.
We infer that the solar wind's Alfvénicity can statistically map near-Earth observations to their source regions because the Alfvén wave forcing that the solar wind experiences in transit is a consequence of the source region topology.

\end{abstract}


\section{Introduction \label{sec:intro}}

The solar wind is a stream of charged particles that continuously leaves the Sun.
In \citeyear{Biermann1957}, \citet{Biermann1957} first observed it from observations of comet tails.
The following year, \citet{Parker1958a} showed that the solar wind is necessarily supersonic and that its speed transitions from subsonic to supersonic by converting thermal energy to kinetic energy at distances of \Rs[\sim5], where \Rs\ is the Sun's radius.
This distance is often referred to as the ``sonic point''.
This supersonic transition is necessary for the solar wind to obtain the minimum speed required to escape the Sun's gravitational pull \citep{Parker1958a,Meyer-Vernet2007a}.
However, the mechanism that converts the solar plasma's thermal energy into kinetic energy does not provide sufficient energy for the solar wind to reach the asymptotic, fastest non-transient speeds observed near Earth.
Additional energy must be deposited into the solar wind above the sonic point for it to reach the asymptotically fastest, non-transient speeds observed near Earth \citep{Leer1980,Hansteen2012,Holzer1981,Holzer1980a,Johnstone2015}.

Detailed analysis of the solar wind in the inner heliosphere and near Earth show that slow and fast speed wind have distinct properties.
For example, the wave modes carried by the plasma also differ in fast and slow wind \citep{Marsch1990}.
In fast wind, fluctuations in the components of the solar wind velocity and magnetic field are highly correlated, while fluctuations in the plasma density and magnetic field magnitude are minimal \citep{Belcher1969,Belcher1971}, as is typical of Alfvénic fluctuations \citep{Alfven1942,Alfven1943}.
In slow wind, compressive fluctuations are also often observed.
Non-thermal features like helium and heavy ion temperatures in excess of mass proportionality with hydrogen ($\Trat[s][\Hy] > m_s/m_{\Hy}$) and speeds along the magnetic field in excess of hydrogen are often observed in fast wind, but absent in slow wind \citep{Kasper2008,Kasper2017,Tracy2015,Tracy2016,Stakhiv2016,Alterman2018,Berger2011,Klein2021,Verniero2020,Verniero2022,Durovcova2019}.
Fast and slow wind also have different chemical and compositional signatures \citep{vonSteiger2000,Geiss1995,Geiss1995b,Zhao:InSituComposition:Sources,Zhao2022,Xu2014,Fu2017,Fu2015,Ervin2023}.
Remote observations show that changes in these chemical and compositional signatures reflect the magnetic topology of the solar wind's source regions on the Sun \citep{Brooks2015,Baker2023}.
Solar wind with fast speeds is typically associated with source regions that have magnetic fields that are radial and continuously open to the heliosphere \citep{Phillips1994,Geiss1995,Arge2013,Wang1990}.
In contrast, sources of slow wind have more complex magnetic field topologies and are only intermittently open to interplanetary space \citep{Baker2023,Antiochos2011,Crooker2012,Abbo2016,Antonucci2005}.

These differences in solar wind properties and their association with solar wind speed has lead to a general identification of near-Earth slow wind with \vsw[400][<] with intermittently open source regions and \vsw[600][>] with continuously open source regions \citep{Bavassano1991,Schwenn2006,Fu2018,Wind:SWE:ahe:xhel}.
However, this classification by speed is known to be imprecise \citep{DAmicis2021a,Yardley2024}, especially in the intermediate speed range \vsw*[400] to \kms[600].
This imprecision is likely because the maximum speed attainable by solar wind from intermittently open regions is greater than the minimum speed of solar wind born in continuously open regions \citep{Wind:SWE:ahe:xhel}, which is a proposed source of the Alfvénic slow wind \citep[e.g.][]{DAmicis2015}.
The Alfvénic slow wind is a subset of solar wind with slow speeds but other properties that are more similar to fast wind from continuously open source regions than slow wind from intermittently open source regions \citep{DAmicis2021a,DAmicis2021,DAmicis2018,Damicis2016,DAmicis2015,Yardley2024}.

The minimum solar wind speed observed in the near-Sun environment follows a Parker-like trend \citep{Raouafi2023}.
These observations also display high levels of Alfvénic fluctuations, irrespective of solar wind speed \citep{Raouafi2023}.
The processes accelerating the solar wind during transit through interplanetary space to speeds in excess of Parker's prediction differ based on solar wind's source region.
Solar wind originating in continuously open source regions is accelerated by the deposition of energy carried by Alfvén waves \citep{Rivera2024,Halekas2023,Perez2021}, while solar wind born in intermittently open source regions is accelerated by thermal and electric pressure gradients \citep{Rivera2025,Halekas2023,Halekas2022}.
This in situ acceleration can be described by the isopoly solar wind model \citep{Dakeyo2022} that includes Alfvén wave forcing \citep{Shi2022} in which the solar wind follows a Parker-like, isotropic expansion in the corona out to some distance $r_0$ and, at heights $r > r_0$, the acceleration decreases and the solar wind follows a polytropic expansion.
Depending on the source region and therefore relevant in situ acceleration mechanism \citep{Rivera2024,Rivera2025}, this critical distance $r_0$ varies between \Rs[6.5] and \Rs[11.5] and the polytropic index varies between $\gamma = 1.3$ and $1.41$.

\PlotHist
The solar wind's total energy flux (\W)  is independent of solar wind speed (\vsw) and solar activity \citep{Lechat2012}.
It consists of the flux of kinetic (\Wk), thermal ($W_H$), and gravitational potential energy ($W_G$) along with the electric coupling between ionized hydrogen and electrons ($W_e$), the flow of energy associated with waves in the plasma (\WA), and the heat flux ($Q$) \citep{Chandran2011,Telloni2023}, the latter of which is negligible far above the sonic point \citep{Telloni2023}.
As show by Parker Solar Probe (Probe) \citep{PSP} and Solar Orbiter (Orbiter) \citep{SO:A,SO:B}, \Wk\ must become an increasing fraction of \W\ with increasing distance from the Sun due to the solar wind's in transit acceleration \citep{Liu2021c,Telloni2023,Rivera2024,Rivera2025}.
We show that the radial scaling of the kinetic energy flux derived from early Probe results \citep{Liu2021c} is sufficient to account for the fast wind peak in the distribution of solar wind speeds observed near Earth.
Comparing this with the results of the isopoly model as applied to the conjunction between Probe and Orbiter in February 2022 \citep{Rivera2024}, we infer that the solar wind acceleration due to Alfvén wave forcing observed below the orbit of Venus continues out to Earth.
Comparing these observations to the range of speeds over which solar wind observations from continuously open source regions become the dominant source of near-Earth solar wind \citep{Wind:SWE:ahe:xhel,Wind:SWE:ahe:dnn}, the range of speeds solar wind from continuously open source regions would reach near Earth in the absence of Alfvén wave forcing, and the impacts of magnetic field topology on solar wind acceleration in the middle corona, we infer that solar wind Alfvénicity is a better predictor of solar wind source region than speed because it is driven by the magnetic field topology of the source region.

\section{Observations \label{sec:obs}}

We use observations of ionized hydrogen (protons) and fully ionized helium (alpha particles) from the \emph{Wind} spacecraft Faraday cups \citep{Wind:SWE} collected near-Earth and outside of Earth's magnetosphere 
for which the fitting algorithms return physically meaningful velocities, densities, and temperatures of hydrogen and helium \citep{Wind:SWE:bimax}.
We select data from the declining phase 23 or 24 \citep{DAmicis2021} when continuously open source regions are confined to the Sun's polar region and intermittently open source regions are confined to the Sun's equatorial region \citep{McComas2008,vonSteiger2000,Schwenn2006}, leading to distinct slow and fast wind peaks in the solar wind speed distribution \citep{Bavassano1991}.
This results in more than 0.75 million distinct solar wind observations of both hydrogen and helium collected over 1828 days.

\section{Analysis \label{sec:methods}}

\TblFit
\cref{fig:hist} plots the distribution of the solar wind speed (\vsw) from the time period selected in \kms[5.38] wide bins.
For visual clarity, this histogram has been smoothed \citep{SavgolFilter,scipy} and values below $10^{-4}$ are not plotted.
This has no impact on our results nor interpretation.

To estimate the location of the fast and slow wind components of the near-Earth \vsw\ distribution, we select data with speeds \vsw[400][<] and \vsw[600][>] and fit each peak with a Gaussian distribution $g(x) = A \cdot \exp \left[-\frac{1}{2}\left(\frac{x - \mu}{\sigma}\right)^2\right]$ with amplitude $A$, mean $\mu$, and standard deviation $\sigma$.
This data selection yields approximately 400,000 distinct slow wind and just over 73,000 fast wind observations.
A plot with the histogram and fits is included in the supplemental material.
\cref{tbl:fits} summarizes the derived fit parameters.
The speeds of the peaks are \vslow[355 \pm 44] and \vfast[622 \pm 58], where we quote the standard deviation of the distribution instead of the fit parameter uncertainty.
These ranges are highlighted with turquoise and labeled.
As slow and fast wind are broadly considered to originate in different classes of source region, there must be a speed below which intermittently open source region is the dominant source of solar wind in the distribution and above with continuously open source regions are the dominant source. 
We estimate this speed \vi[484 \pm 34] as the range of speeds over which the Gaussians fit to both the fast and slow subsets intersect, when the uncertainties on all fit parameters are propagated.
\cref{fig:hist} indicates \vi\ in pink.
\cref{tbl:v} summarizes all speeds derived and quoted in this work.

\TblVel
\citet{Liu2021c} show that near-Sun observations of the solar wind's total and kinetic energy flux obey power laws of the form 
\eqRadialScaling
\cref{tbl:scaling} summarizes the coefficients $c$ and exponent $b$ derived by \citet{Liu2021c}.
Combining both power laws yields a near-Earth kinetic energy flux estimate \Wk[0.64 \pm 0.02].

The kinetic energy flux is given by \cref{eq:Wk}, where \m[s]\ is the mass density for a given species and \vel<s> is its speed.
\eqWk
We calculate the total \Wk\ from near-Earth observations considering helium and hydrogen.
If we assume the electrons, hydrogen, and helium move at the same speed, quasi-neutrality implies that the electron contribution to the kinetic energy flux is on the order of $0.05\%$ and therefore negligible.

\cref{fig:KE} plots the frequency of near-Earth solar wind observations as a function of solar wind speed and kinetic energy flux.
Because slow solar wind is observed markedly more frequently near-Earth than fast wind, each column has been normalized to its maximum value so that this sampling artifact is removed.
To extract the trend of \Wk\ as a function of \vsw, we bin the data in \kms[10] wide speed intervals and fit each column with a log-normal distribution.
\cref{fig:log-normal-fit} is an example of one such fit.
We use a log-normal distribution because the distribution tails are asymmetric and enhanced on the side of the distribution corresponding to higher \Wk.
The resulting fit parameters are plotted in black.
The central trend (dash-dotted line) is the peak from these 1D fits.
The shaded region bounded by thin black lines corresponds to their widths.
Only distributions up to \kms[740] are well-described by a log-normal distribution.
This is unsurprising because we select data from solar minima and these speeds correspond to the region of \cref{fig:hist} where the PDF of \vsw\ decrease markedly.
The horizontal orange region is the range of near-Earth values predicted above using near-Sun observations \citep{Liu2021c}.
As the orange region intersects the solid black lines below \kms[740], this has no impact on our conclusions.

\TblScaling
To calculate the speeds at which the near-Sun estimate of the near-Earth kinetic energy flux ($W_\mathrm{est}$) and the observed near-Earth kinetic energy flux ($W_\mathrm{obs}$) intersect, we have constructed a spline representation of $W_\mathrm{obs}$ (dash-dotted line in \cref{fig:KE}) and calculated the speeds at which it intersects $W_\mathrm{est}$, \vWk[642].
\cref{fig:KE} shows that the minimum speed at which these two values for \Wk\ intersect corresponds to the upper bound on the $W_\mathrm{obs}$ trend and the minimum of $W_\mathrm{est}$.
The maximum corresponds to the intersection of the maximum value $W_\mathrm{est}$ and the lower bound on $W_\mathrm{obs}$.
Applying the same spline method to these intersections as above, we determine that this range of speeds corresponds to \vsw*[557] to \kms[700].
\cref{fig:hist} plots \vWk\ as a vertical orange bar and highlights the full range of values in orange.

\PlotKeFluxCombined

The difference between \vfast[622]\ and the center of the \vWk\ range of values is \kms[20].
The maximum value for \vi\ is \kms[518], which is \kms[39] slower than the minimum of \vWk\ and almost twice as large as the difference between the central \vfast\ and \vWk\ values.
In other words, the near-Earth fast solar wind speeds predicted from near-Sun observations of \Wk\ correspond well with \vfast\ observed near-Earth.
Although this range of predicted near-Earth speeds is wider than the \vfast\ peak, it is well separated from the range of speeds \vi\ for which slow and fast solar wind are both observed to have a non-trivial contribution in the near-Earth environment.

Assuming the conjunction between Probe and Orbiter that was described by the isopoly model with Alfvén wave forcing \citep{Rivera2024} is representative of the evolution in transit of solar wind from magnetically open sources, we use the range of values that account for the fit uncertainties on the Alfvén wave forcing profile along with the uncertainties on $\gamma$ to calculate that the corresponding range of expected speeds near-Earth (\Rs[212]) for this event is \vIPW*[620] to \kms[671] with a central value of \kms[645].
The slowest \vIPW\ corresponds to the largest $\gamma$ and weakest Alfvén wave forcing profile, while the fastest \vIPW\ corresponds to the smallest $\gamma$ and the strongest forcing profile.
This range of \vIPW\ falls within the range of speeds given by \vfast\ and \vWk.
\cref{fig:hist} labels this range of speeds \vIPW\ and highlights it in green.

\PlotLogNormalFit

The temperature in the isothermal layer consistent with \citepossessive{Rivera2024} observations and the isopoly that includes Alfvén wave forcing is \Temp_o_[1.7].
To determine the range of temperatures in the isothermal layer corresponding to \vfast, we minimize the absolute value of the difference between the near-Earth speed predicted form the isopoly model and the fast wind peak $\left|\ve\!\left(T_o,f_0,\gamma\right) - \vfast\right|$ over the ranges $\gamma = 1.41 \pm 0.02$ and $f_0 = 0.012 \pm 0.002$ in units $GM_\odot/R_\odot^2$ from \citet{Rivera2024}, where $f_0$ sets the strength of the Alfvén wave forcing profile.
Using \vfast[622 \pm 58], the range of isothermal layer temperatures is \Temp*_o_[1.34] to \MK[1.85] with a central value of \Temp_o_[1.6], which are consistent with UVCS observations \citep{Cranmer2020} and exceed the estimates of \Temp_o_ from exospheric theory using near Sun electron observations \citep{Halekas2022}. 
The corresponding Alfvén wave forcing strengths are $f_0 = \pten[1.2]{-2}$, \pten[1.22]{-2}, and \pten[1.18]{-2}.
The polytropic index is $\gamma=1.41$ in all cases.
Under the isopoly model, the near-Earth speeds corresponding to these \Temp_o_ in the absence of Alfvén wave forcing is \vIP*[434] to \kms[554] with a central value of \kms[497].
\cref{fig:hist} labels this range of speeds \vIP\ and highlights it in green.

The helium abundance is given by helium-to-hydrogen number density ratio in units of percent $\ahe = 100 \times \n[\He]/\n[\Hy]$.
In the slow wind, \ahe\ is highly variable below $4.19\%$, and $\grad_\vsw_[\ahe] > 0$ \citep{Wind:SWE:ahe:xhel,ACE:SWICS:FSTransition}.
In fast wind, the helium abundance is fixed at \ahe[4.19], which is 49\% of the photospheric abundance, and the gradient is flat \citep{Wind:SWE:ahe:xhel,ACE:SWICS:FSTransition}.
This change in \grad_\vsw_[\ahe] is one composition signature suggesting that slow solar wind originates in intermittently open source regions, while fast wind originates in continuously open source regions \citep[Section 4.3 and references therein]{Wind:SWE:ahe:xhel}.
Fitting a bilinear function of $\fcn{\ahe}{\vsw}$, \citet{Wind:SWE:ahe:xhel} define the speed and abundance at which this change of gradient occurs as the saturation point with saturation speed \vs\ and saturation abundance \As.
They show that \vs\ decreases as the Alfvénicity increases, while \As\ increases with increasing Alfvénicity.
Considering the solar wind Alfvénicity as a proxy for the magnetic field topology above the sonic point, \citet{Wind:SWE:ahe:xhel} argue that the dependence of \vs\ and \As\ on Alfvénicity are consistent with the minimum \vsw\ observed at \au[1] for solar wind in continuously open source regions is less than the maximum \vsw\ for solar wind observed at \au[1] and born in intermittently open source regions.
\citet{Wind:SWE:ahe:dnn} show that the unexpected presence of compressive fluctuations in fast wind modifies \vs\ and \As.
Accounting for these effects, the range of speeds over which solar wind from continuously open sources become a significant component of the near-Earth solar wind speed distribution is \vsigma*[399] to \kms[436], which is just faster than the fastest \vslow\ and overlaps with \vIP\ by \kms[2] \citep{Wind:SWE:ahe:dnn}.
\cref{fig:hist} plots this range of speeds in pink and labels it \vsigma, as $\sigma$ is a common symbol for the Alfvénicity.

\section{Discussion \label{sec:disc}}

Recent observations suggest that differences in the speed of solar wind form intermittently and continuously open source regions begin to develop in the middle corona at heights above the Sun's surface \Rs[2 - 3.5] \citep{Ngampoopun2025}.
In continuously open field regions, the speed is highly correlated with magnetic field geometry, while it is not in intermittently open field regions \citep{Ngampoopun2025,Wang1990}.
Above the middle corona, the minimum speed of near-Sun solar wind 
is consistent with Parker's isothermal solar wind model \citep{Parker1958a,Raouafi2023}.
During propagation through interplanetary space, solar wind from intermittently open source regions is accelerated by thermal pressure gradients \citep{Rivera2025}.
In contrast, solar wind from continuously open source regions is accelerated by both thermal pressure gradients and Alfvén wave forcing \citep{Rivera2024,Rivera2025}.
Theory and observations suggest that the physical mechanisms generating Alfvén waves and producing the energy necessary to accelerate the solar wind during interplanetary propagation are common to  intermittently and continuously open source regions and they are sufficient to accelerate the solar wind to the speeds \vfast\ observed near Earth \citep{Fisk2001,Fisk2005,Fisk2020,Zank2020,Bale2023,Raouafi2023,Wyper2022}.

The solar wind's Alfvénicity is a more reliable predictor of solar wind source region than speed \citep{Tu1992a,DAmicis2021}.
The Alfvénicity quantifies whether the dominant wave modes in the plasma correspond to high correlations between the velocity and magnetic field components in the absence of density compressions and fluctuations in the magnetic field magnitude \citep{Belcher1969,Belcher1971,DAmicis2021,DAmicis2021a,LR:turbulence}.
Such signatures are considered to be indicative of the presence of Alfvén waves \citep{Alfven1942,Alfven1943}.
Fast solar wind from magnetically open sources has a high Alfvénicity.
Solar wind from magnetically closed sources does not display a preferred Alfvénicity, which means that it can carry a wide range of fluctuations and waves \citep{DAmicis2021a,Wind:SWE:ahe:xhel}.
The Alfvénicity of fluctuations carried by solar wind from intermittently open sources decays during solar wind propagation through interplanetary space, while the Alfvénicity of solar wind from continuously open source regions does not \citep{Marsch1990}.
Large scale motion of the Sun's corona are likely a key source of solar wind Alfvén waves \citep{PSP:SWEAP:1,Fisk2001,Fisk2005} and can drive the solar wind's Alfvénicity.
When the solar wind speed is larger than the Alfvén speed, this is no longer the case.
This height is the Alfvén radius, nominally \Rs[\sim20] \citep{PSP:SWEAP:Ra}, well above the middle corona where differences in solar wind speed based on the magnetic topology of source regions begin to develop.

We have estimated the range of near-Earth fast wind speeds \vfast[622 \pm 58] and slow wind speeds \vslow[355 \pm 44]  using the distribution of \vsw\ observed by the Wind spacecraft during the declining phase of solar activity when continuously and intermittently open source regions are heliographically stratified.
\citet{Wind:SWE:ahe:xhel,Wind:SWE:ahe:dnn} suggest that the minimum speed at which solar wind from continuously open source regions becomes a non-trivial component of the distribution of \vsw\ observed near Earth is \vsigma*[399] to \kms[436].
To naively estimate the range of slowest speeds over which solar wind from continuously open source regions becomes the primary component of near-Earth observations, we have calculated the speed \vi[484 \pm 34], the range of speeds over which Gaussians fit to the slow and fast wind distributions intersect.

Empirically derived radial scalings of the solar wind's kinetic energy flux \citep{Liu2021c} lead to a predicted near-Earth speed \vWk*[557] to \kms[700].
\vWk\ spans the range of fast wind speeds \vfast\ and exceeds \vi.
This suggests that the radial evolution of the solar wind's kinetic energy flux is sufficient to account for the statistical range of fast solar wind speeds observed near Earth.

The range of near-Earth solar wind speeds predicted by a solar wind model that includes both thermal pressure gradients and Alfvén wave forcing \citep{Rivera2024} is \vIPW*[620] to \kms[671], which overlaps with the central values of both \vfast\ and \vWk, but is significantly narrower than both.
Under this same model, near-Earth solar wind with speeds \vfast\ would only reach speeds \vIP*[434] to \kms[554] if it was not accelerated by Alfvén wave forcing during transit through interplanetary space.
The range of isothermal temperatures used to derive \vIP\ varies by 38\%, while the strength of Alfvén wave forcing profiles varies by 3\% and the polytropic index remains fixed at $\gamma=1.41$.
This suggests that the range of near-Earth fast wind speeds is sensitive to the temperature boundary conditions, while insensitive to the overall strength of the forcing profile and the in situ heating.
The speed range \vIP\ is at least \kms[35] faster than the maximum \vslow, spans \vi, is \kms[10] slower than the slowest \vfast, and overlaps \vsigma\ by at most \kms[2], which is less than 50\% the width of a single bin in the PDF of \vsw\ in \cref{fig:hist}.
This suggests that Alfvén wave forcing of the solar wind continues out beyond the orbit of Venus, consistent with observations of the decay of Alfvénic structures (e.g. switchbacks) as the solar wind propagates through interplanetary space \citep{Tenerani2021,Rasca2021}.

In this work, we have de facto declared that speed differences or overlaps on the order of \kms[\lesssim 10] near Earth are insignificant.
The size of the histogram bins used to generate \cref{fig:hist} is \kms[5.38].
The full range of non-transient \vsw\ observed at \Rs[212] is \vsw*[200] to \kms[800].
A speed range of \kms[10] is approximately 1.7\% of this \kms[600] range.
The maximum range of speeds covered by \vslow\ to \vWk\ is \vsw*[311] to \kms[700].
A speed range of \kms[10] is approximately 2.6\% of this \kms[389] span.
The observations from \emph{Wind} near Earth consist of two 913 day intervals collected during the declining phase of solar activity, separated by at least 8.5 years and ending in 2018.
The speed estimate \vWk\ is derived from near-Sun observations collected over four different 12-day intervals spanning 2018 to 2022 \citep{Liu2021c}, which spans the end of the declining phase of solar cycle 24 and minimum 25.
The model proposed by \citet{Rivera2024} is developed using 1hr and 40min of Probe observations that correspond to 10 hrs of Orbiter observations collected over the course of two days in 2022, during the rising phase of solar cycle 25.
Given that this work compares four different sets of observations, collected over four different ranges of time, at different phase of solar activity, and at different radial distances, we have performed our analysis using the widest possible ranges of values that account for the uncertainties on all derived speeds and model parameters.
Only using the central or mean values would only reduce the ranges of speeds and improve our results.
As such, a speed difference or overlap on the order of 2 histogram bin widths 
seems negligible.
As such, we infer that the reason that Alfvénicity is a reliable predictor of source region is because solar wind from continuously open source regions is accelerated in transit by Alfvén wave forcing, while solar wind from intermittently open source regions is not, and the presence or absence of Alfvén wave forcing during interplanetary propagation is a consequence of the magnetic field topology at the source region.

\begin{acknowledgements}
The author thanks the referee for their helpful suggestions.
The author thanks M.~M.~Martinvoić, M.~Liedel, C.~M.~G.~Kevorkian, and A.~Szabo for helpful comments on the manuscript and S.~A.~Livi and R.~D'Amicis for useful discussions.
B.L.A. is funded by NASA 
grants 80NSSC22K0645 (LWS/TM), 80NSSC22K1011 (LWS), and 80NSSC20K1844.
B.L.A. also acknowledges NASA's support of the Parker Solar Probe and Solar Orbiter missions at NASA/GSFC.
\end{acknowledgements}

%
%
%
%
%
%
%
%
%
%

\bibliography{Zotero.bib}{}
\bibliographystyle{aasjournal}

\end{document}